# Shaping the future of Global Interferometric Arrays

## Imaging Strong Gravity and Magnetic Fields

**Authors:**

**Venkatessh Ramakrishnan**[1] *Signal Processing Research Centre, Tampere University, Finland*
**Violette Impellizzeri** *ASTRON, Netherlands Institute for Radio Astronomy*

***Chi-Kwan Chan*** *(University of Arizona, USA),* ***Mariafelicia De Laurentis*** *(UNINA, Italy),* ***Thomas Krichbaum*** *(MPIfR, Germany),* ***Andrei Lobanov*** *(MPIfR, Germany),* ***Laurent Loinard*** *(UNAM, Mexico),* ***Freek Roelofs*** *(Radboud University, The Netherlands),* ***Eduardo Ros*** *(MPIfR, Germany)* ***Hannah R. Stacey*** *(ESO, Germany)*

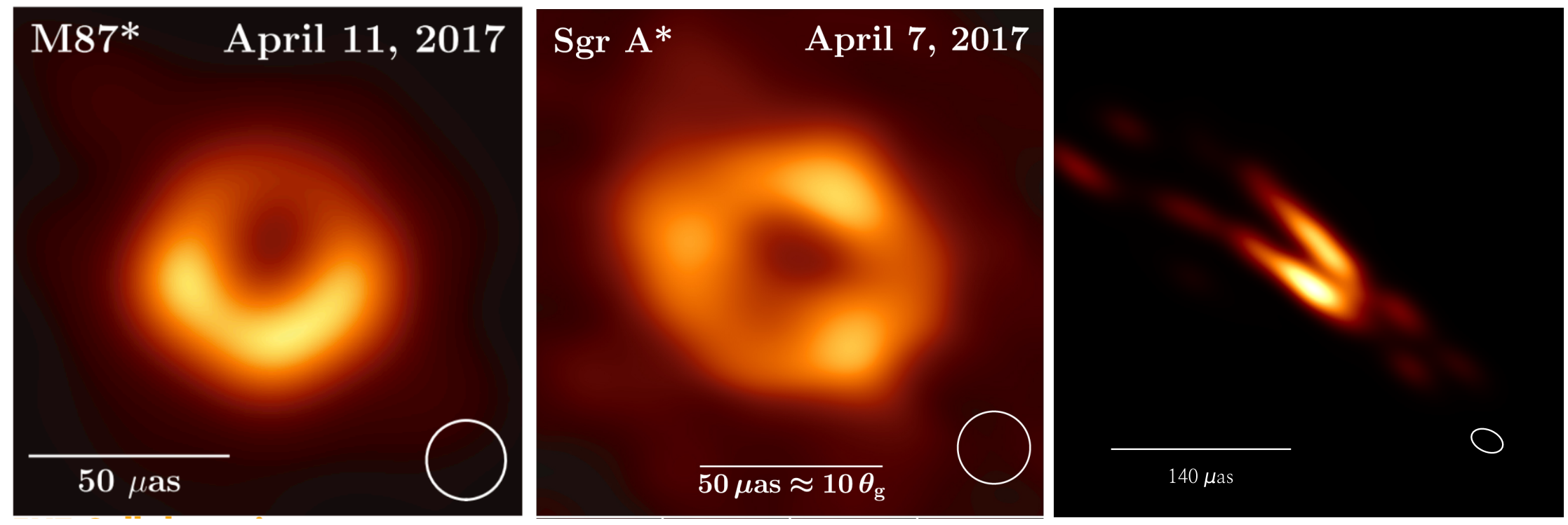

Images from 2017 EHT+ALMA observations of M87* (left), Sgr A* (centre) and Centaurus A (right).

[1] venkatessh.ramakrishnan@tuni.fi

**Abstract:** The observational validation of General Relativity (GR) has been propelled in recent years by recent breakthroughs in Very Long Baseline Interferometry (VLBI) augmented by ALMA. We explore ALMA2040 opportunities to transform these studies through greatly improved sensitivity and a multi-frequency approach. The focus will be on placing most stringent constraints on GR and alternative theories in the strong-gravity regime, and on understanding the formation and launching of relativistic jets.

# 1 Scientific Context and Open Questions

We are in an era where high-precision cosmology is pushing the limits of existing observational infrastructure both in sensitivity and angular resolution. Very Long Baseline Interferometry (VLBI) plays a vital role in this context by delivering the highest angular resolution at radio to submillimetre wavelengths. Here, we focus on the (sub-)millimetre regime, where ALMA still plays a dominant role. Over the past decade, VLBI has undergone major advances with the establishment of the Event Horizon Telescope (EHT) network, driven by years of pioneering international collaboration (Doeleman 2008; Krichbaum et al. 2008). The decisive breakthrough, however, came with the inclusion of ALMA in the EHT and the Global mmVLBI Array (GMVA). Building on the scientific outcomes of these observations, we outline the following science cases, which primarily highlight studies of active galactic nuclei (AGN) and their associated phenomena.

## 1.1 High-precision laboratory for testing gravitational physics

The EHT+ALMA results on M87$^*$ (EHTC et al. 2019) and Sgr A$^*$ (EHTC et al. 2022) directly and almost unequivocally provide evidence for the existence of at least two supermassive black holes (SMBHs). Following these results, the focus has now shifted to refining the angular resolution and image fidelity by at least a factor $\times 3$ or higher in Sgr A$^*$ and M87$^*$ to sample the photon ring and to sharpen measurements of the central brightness depression and ring morphology.

The expectation for the appearance of a black hole shadow is firmly grounded in the assumption that Einstein's theory of General Relativity (GR) provides the correct description of gravity. While GR remains the most successful theory of gravity to date, it is not the only viable framework. For instance, observational evidence for dark matter has motivated alternative theories involving scalar fields. Likewise, the discovery of a dominant, yet unknown, form of energy driving the accelerated expansion of the Universe (*dark energy*) has led to the development of a broad class of modified gravity theories. These frameworks can naturally account for the cosmological expansion and, therefore, pose a direct challenge to the validity of GR (Mizuno et al. 2018). Using high-angular resolution and improved sensitivity of VLBI+ALMA, the goal of this science case is to investigate the following question.

- How accurately can we determine the mass, spin, and quadrupole moment of nearby SMBHs, and do they satisfy the predictions of the Kerr metric?
- Do multi-frequency VLBI observations reveal deviations from GR strong-field predictions, such as shadow distortions or photon-ring structure?
- Can we detect signatures of frame dragging or disk warping that directly probe spacetime geometry near the SMBH?
- What observational precision is required to distinguish GR from modified-gravity scenarios in the strong-field regime?

In this roadmap, the most anticipated development is the simultaneous multi-band observations at frequencies 86/230/345/690 GHz, which will increase the coherence time for 690 GHz observations and superior time sampling across all frequencies. The 345 and 690 GHz observations will probe regions closer to the black hole, since the optical depth of the M87$^*$ and Sgr A$^*$ accretion disks are lower, while the effects of interstellar scattering in Sgr A$^*$ observations are reduced (Johnson & Narayan 2016). Taken together, the jump to 345 and 690 GHz turns both primary targets into more precise laboratories for studies of the black hole spacetime.

### 1.2 Revealing the innermost accretion flows around supermassive black holes

Although the first two images of an event horizon are a breakthrough discovery, significantly more work will be required to provide precision tests of GR (or alternatives) and to extrapolate the results to the general and diverse population of all SMBHs. To this end, we use the SMBH catalogue of Ramakrishnan et al. (2023), which identifies at least a dozen nearby systems whose accretion flows can be imaged. Extending observations to Band 9 (690 GHz) alone increases this number to $\sim$30, enabling resolution of the inner photon ring in the nearest AGN. The addition of a space baseline would further expand the sample to $\sim$50. *Even an increase to $\sim$30 represents a fundamental advance to our understanding of AGN accretion.*

The primary driver for addressing these questions is to understand the friction generated around black holes, which causes the formation of accretion disks. In this endeavour, we aim to use ALMA2040 to directly address the following questions.

- What are the magnetisation, electron temperature distribution, and optical-depth structure of the innermost accretion flow?
- What sets the radial extent and geometry of the emission region close to the event horizon?
- Which physical mechanisms trigger flares, and how do they connect to turbulent MHD processes near the SMBH?
- How does the accretion-flow variability evolve across different timescales, and what does this reveal about plasma dynamics?
- Can VLBI constrain angular-momentum transport and the role of magnetic stresses in shaping the inflow?

A crucial element for characterising inner accretion flows is multi-frequency polarimetry. Simultaneous 86/230/345 GHz VLBI with ALMA2040 will allow detailed Faraday rotation studies, constraints on the magnetization of the plasma, and dissimination between electron–ion and pair-dominated flows. High-frequency circular and linear polarization measurements with a dynamic range of 1000:1 will provide direct insight into the geometry and strength of magnetic fields, enabling stringent tests of RIAF and jet–disk coupling models.

The limiting factor in addressing these goals with the current generation of EHT+ALMA is its sensitivity. Since a majority of the targets from the sample are relatively faint, it is challenging to obtain strong fringes on continental baselines in addition to phasing up ALMA. ALMA in its current state provides a sensitivity to image targets as faint as 50–100 mJy. While this has been addressed in the form of passive phasing, a factor of 10$\times$ increase in the sensitivity of ALMA will allow phasing to be done more easily, reaching fainter science and calibrator targets.

### 1.3 Ultra high-resolution view of the Jet launching regions

The strong (accreted) magnetic fields close to the event horizon of SMBHs are posited to be responsible for launching bipolar relativistic plasma jets. Although several mechanisms have been proposed for jet launching (e.g. Blandford & Znajek 1977; Blandford & Königl 1979; Blandford & Payne 1982), with energy extraction either from the accretion flow or from black hole spin, the physics of jet launching remains an active area of research and is primarily limited by the paucity of observational data. The jets are expected to be more collimated right from their base when driven by the black hole, while in the case of disk-driven jets, it is expected to have a wider opening angle (e.g. Walker et al. 2018; Kim et al. 2019).

Numerical simulations of jet formation invoke strong magnetic fields. This, when complemented with the motions and polarisation structure in jets, can provide clues to the acceleration process, jet composition and their energetics. Here, we look forward to addressing the following issues.

- What is the process responsible for launching powerful collimated jets with black holes of varying spins?
- What drives these relativistic jets, is it the black hole or the accretion disk?
- What is the significance of jets in galaxies with bright and faint radio jets?
- What is the transverse structure and collimation profile of the jet?

These questions require high dynamic range imaging at intermediate baselines (tens to hundreds of microarcseconds), full-Stokes VLBI at 230/345/690 GHz to map magnetic fields and Faraday-rotation gradients, and participation in extended EHT campaigns (months with sub-week cadence) to produce genuine jet "movies" rather than isolated snapshots.

## 2 Technical Requirements

To address the above-mentioned science cases, we require the next-generation (sub-) mm interferometer to have a sensitivity providing a dynamic range of the order of 1000:1 in VLBI observations. Such highly sensitive observations can only be obtained by increasing the bandwidth to 128 GHz (a factor $\times 4$ better than the planned wideband sensitivity upgrade) *at least* at 230 and 345 GHz. Our science goals would also strongly benefit from Band 9 (690 GHz) observations, allowing to image optically thin emission at even higher angular resolution. Another important requirement is simultaneous multi-frequency capability at frequencies 86/230/345/690 GHz, which will be a game-changer allowing to increase coherence time at the highest frequencies (Dodson et al. 2023; Zhao et al. 2025). The objective is to have a homogeneous VLBI array since all current and future facilities that are part of the EHT and GMVA networks are planned to converge to a higher bandwidths by the next decade, *and* to increase in the number of available high frequency bands. The choices made under ALMA2040 will be pivotal in shaping the future dynamics of global arrays. Finally, increasing the number of antennas by a factor $\times 4$ (assuming 15 m dishes), we can synthesize a 200-m hybrid dish which will dramatically increase the sensitivity of our global observations and allow us to detect 10–20% fainter black hole rings.